\documentstyle[12pt,amssymb,epsfig,color]{article}

\begin{document}

\def \cl {{\cal L}}
\def \be {\begin{equation}}
\def \ee {\end{equation}}
\def \bea {\begin{eqnarray}}
       \def \eea {\end{eqnarray}}
\definecolor{agr}{rgb}{0.0,0.4,0.0}
\definecolor{apu}{rgb}{0.3,0.2,0.6}
\definecolor{dre}{rgb}{0.6,0.0,0.1}
\def \bi {\bibitem}
\def \ci {\cite}
\def \e {{\rm e}}
\def \o {\omega}
\def \a {\alpha}
\def \n {\nu}
\def \cf {{\cal F}}
\def \L {\Lambda}
\def \x {\xi}
\def \g {\gamma}
\def \del {\partial}
\def \pt {{\rm PT}}
\def \ep{\epsilon}
\def \eps {\varepsilon}\newcommand\re[1]{(\ref{#1})}%
\def \lab #1 {\label{#1} \mbox{\# ${#1}$}}
\def \lab #1 {\label{#1}}
\def \p {\pi}\def \m {\mu}
\def \cs {{\cal S}}
\def \as {{\alpha_s}}
\def \txc {\textcolor}
\def \top {\vbox{\vskip 0.25 true in}}
\def \Top {\vbox{\vskip 0.5 true in}}


\begin{flushright}
BNL-NT-04/28 \\
RBRC-432  \\
YITP-SB-04-44 \\
\today
\end{flushright}

\vspace*{20mm}

\begin{center}

{\LARGE {\bf Recoil and Power Corrections

\bigskip
\bigskip

in High-$x_T$ Direct-Photon Production }}

\par\vspace*{15mm}\par

{\large George Sterman$^a$ and Werner Vogelsang$^b$}

\bigskip

{\em $^a$C.N.\ Yang Institute for Theoretical Physics,
Stony Brook University \\
Stony Brook, New York 11794 -- 3840, U.S.A.} \\

\bigskip

{\em $^b$RIKEN-BNL Research Center and Nuclear Theory, \\
Brookhaven National Laboratory,
Upton, NY 11973, U.S.A.}

\end{center}
\vspace*{15mm}

\begin{abstract}
\noindent
We study a class of nonperturbative corrections
to single-inclusive photon cross sections at
measured transverse momentum $p_T$, in the large-$x_T$ limit.
We develop an extension of the joint (threshold and
transverse momentum) resummation  formalism,
appropriate for large $x_T$, in which there are no kinematic
singularities associated with recoil, and
for which matching to fixed order and to threshold
resummation at next-to-leading logarithm (NLL) is straightforward.
Beyond NLL, we find contributions that can be attributed to recoil
from initial state radiation. Associated
power corrections occur as inverse powers of $p_T^2$ and are
identified from the infrared structure of integrals over the
running coupling. They have significant energy dependence and
decrease from typical fixed-target to collider energies.
Energy conservation, which is incorporated into joint
resummation, moderates the effects of perturbative recoil
and power corrections for large $x_T$.

\end{abstract}


\newpage
\section{Introduction}

To study the interplay of
perturbative and nonperturbative dynamics in
processes involving hadronic states it is natural
to begin with observables whose perturbative analysis is
well understood.  For certain observables,
perturbation theory not only provides predictions
at leading power in a large momentum scale,  but also
characterizes  power corrections in that scale.
This can come about, for example, through
nonconvergent perturbative expansions that exhibit sensitivity
to the strong-coupling and/or vacuum structure of the theory
\cite{irrpc}.

Relying on perturbative resummations, this approach has had
phenomenological successes in the description of a variety of inclusive
and semi-inclusive cross sections.  These include average and
differential event shapes, primarily but
not exclusively in ${\rm e^+e^-}$ annihilation
[2-9],
and electroweak annihilation cross sections at
measured transverse momentum
[10-13].
The value of the event
shape or lepton pair transverse momentum provides
a second scale in the cross section, and varying this scale
changes the relative importance of
perturbative and nonperturbative dynamics.
Thus, the transition between perturbative
and nonperturbative QCD is
in principle available for study in these observables.

In this paper, we adopt this general philosophy
and employ the joint resummation \cite{lae00b,li99} of threshold
\cite{thresum} and
transverse momentum \cite{ktresum1} enhancements  to
study  power corrections in the
hard-scattering scale $p_T$ for single-particle inclusive (1PI) cross
sections in the large $x_T^2=4p_T^2/S$ region.
Using direct photon production as an example, we will show
that these corrections exhibit significant $x_T$ dependence,
which moderates both perturbative and nonperturbative recoil
at large $x_T$ compared to
estimates based on transverse momentum resummation
alone.  These conclusions are made possible by a
simplification of the joint resummation formalism
that is specific to the $x_T \rightarrow 1$ limit.

Direct photon production was originally envisioned
as a relatively straightforward process with
which to test fixed-order perturbative
calculations and to determine the
gluon distribution \cite{fritzsch,owe87,dgamnlo}.
The extensive data on direct photon production
[20-24], however,
has turned out to be more complex than was perhaps expected.
Presumably for this reason it
has inspired varied theoretical and phenomenological studies
[25-37].
Nevertheless, for this benchmark process important
questions remain unresolved.
In particular, it has been argued that
fixed target data for direct photon production in the lower $p_T$
range (roughly below 5 GeV) are difficult to reconcile with
collinear-factorized NLO cross sections \cite{apa99,aur00}.
Additionally, threshold resummation \cite{1pith,cmnov,kido}
appears to explain the data only for larger $p_T$.

This difficulty has motivated the use of
$k_T$-dependent, or unintegrated, parton distributions
combined with recoiling partonic $2\rightarrow 2$ subprocesses
\cite{kim99}.
Information on the partonic transverse momenta
in such distributions may
come from  resummed perturbation theory \cite{ktresum1,ktresum2,kim99},
and/or from comparisons to data \cite{bro02,apa99,almur,pipairs},
including Drell-Yan, photon and hadron pair cross sections. Probably the
simplest approach is to assume a Gaussian
dependence $\exp[-k_T^2/\langle k_T^2\rangle]$ \cite{owe87}.
As we review below, perturbative resummations
predict logarithmic $p_T$ dependence for the
parameter $\langle k_T^2\rangle$.  They also imply
that $\langle k_T^2\rangle$ depends on the parton flavor.

The use of unintegrated distributions requires an extension
of collinear factorization \cite{highfact}.
In particular, a technical challenge in the case of light particle
production
is the potential for an
artificial infrared singularity when the total transverse momentum
of the initial state partons is comparable to the observed $p_T$
\cite{owe87}.
One way to avoid this singularity is to impose strong
ordering in transverse momenta, as in \cite{kim99}, a
procedure which requires definition beyond leading logarithm.
A related approach is described 
detail in Ref.\ \cite{fin01}, based on
a specific implementation of $k_T$-resummation.
As presented in \cite{fin01}, however,
fits in this formalism favor fixed
$\langle k_T^2\rangle$, with no indication of
the $p_T$-dependence implied by $k_T$ resummation.
Other studies, however, seem to imply that $\langle k_T^2\rangle$
is $S$-dependent \cite{apa99}.
In summary, it remains unclear how much of what we interpret as
recoil, or parton transverse
momentum, is perturbative and how much nonperturbative.
Here we come back  to this question in the context of
a generalized resummation formalism.

Resummed perturbation theory for 1PI cross sections
was extended  in \cite{lae00b,li99,lae00a} using joint resummation.
Joint resummation systematically combines singular behavior at zero
transverse momentum for initial-state partons
with that at partonic threshold, where the
initial state partonic invariant mass  $\hat{s}=x_ax_bS$ is just large
enough to produce the observed final state.  This
method was applied to Z and  Higgs production in \cite{kul02,kul03},
where no kinematic singularities arise, because the
produced electroweak state is massive and the transverse
momentum of the lepton pair is directly observed.
In \cite{kul02}, some implications for
the specific forms of power corrections were also
pointed out. Although the joint formalism was applied to
high-$p_T$ photon production in \cite{lae00a},
its application was hampered by the same infrared singularity
mentioned above, associated with the production of a massless particle.
As noted in \cite{lae00a}, the complexity of a simultaneous
resummation in transverse momentum and energy above threshold
appears to make impractical a matching of the sort
developed for transverse momentum resummation
alone in \cite{fin01}.

In this paper, we extend this work, and revisit
logarithmic and power corrections to the direct photon cross
section in the joint resummation formalism.
Compared to previous work, however, we use the
kinematics of the large-$x_T$ limit to
reformulate joint resummation, taking into account
recoil effects in the partonic
subprocess while avoiding a kinematic singularity.
The resulting resummed cross section reduces to
threshold resummation at next-to-leading logarithm (NLL) and can be
matched
to finite-order and threshold resummed cross sections
in a straightforward fashion.
Beginning at NNLL, the cross section also includes a contribution
that can be identified as the finite residue left from
the cancellation of the transverse momentum singularities of real
and virtual gluons radiated in the initial state.
Enhancements to the cross section associated with final
state interactions are treated only to leading power in this paper,
and appear in the same manner as in threshold resummation.
In another paper we will argue that the results found here
are not changed qualitatively by these effects.

The parameters
that control power corrections
associated with joint resummation at partonic
threshold are found to be related to parameters familiar from the
transverse momentum distributions in electroweak annihilation.
The power corrections also inherit significant energy dependence.
For large $x_T$, both perturbative recoil and nonperturbative power
corrections to the predictions of threshold resummation are suppressed
by
the phase space restrictions built into joint resummation.  This effect
is
important, however, only for $x_T$ near one, or equivalently for large
values of its conjugate Mellin moment variable $N$.  For smaller $x_T$
or
$N$ of order unity an analysis based on $k_T$ resummation alone may be
appropriate, but should be matched to the results of joint resummation
in
the large $x_T$ region.

We begin Sec.\ 2 with a brief summary of the joint resummation
formula as developed for direct photon cross sections, and exhibit
the kinematic singularity.  In the next subsection, the cross
section is expressed as a double inverse transform.
This is followed by a simple reformulation
that eliminates the kinematic singularity and reduces the
jointly resummed cross section to a single transform
that extends threshold resummation for the direct photon cross
section. The resulting Sudakov exponents of joint resummation are
analyzed in Sec.\ 3, where we identify the form of the recoil and
power corrections to the direct photon cross section
that are implied by joint resummation.
We explore the phenomenology of these corrections
in Sec.\ 4, exhibit the suppression of power
corrections for large $S$ at fixed $p_T$, and
briefly discuss possible subdominant corrections
not directly associated with partonic threshold.
We conclude with a summary, and a brief
discussion of possible implications for an
eventual global treatment of single-photon and
single-hadron cross sections.

\section{Self-consistent Recoil in Joint Resummation}

\subsection{Partonic recoil in direct photon production}

Joint resummation \cite{lae00b,lae00a} is an extension of
threshold  \cite{thresum} and transverse momentum
resummations \cite{ktresum1,ktresum2} that unifies these two formalisms.
So far, at the phenomenological level it has been
applied primarily to the single electroweak boson (mass $Q$) production
cross sections at low transverse momentum, $Q_T\ll
Q$~\cite{kul02,kul03}.
In this case, threshold resummation
is associated with corrections of the form
$[\as^n/(1-z)]\ln^{2n-1}(1-z)$,
with $z=Q^2/\hat{s}$, where $\sqrt{\hat{s}}$ is the invariant mass of
the partonic pair that annihilates into the observed boson. Such
corrections are ``implicit" in the sense that they contribute to
the hadronic cross section only after convolution with the parton
distribution functions, and hence give nonlogarithmic, although
potentially significant, contributions to the cross section.
Singular corrections in $Q_T$, on the other hand, are explicit in
the cross sections themselves, appearing as terms like
${\as^n/Q_T^2}\ln^{2n-1}(Q_T/Q)$ directly for the measured spectrum.

For single-particle inclusive cross sections such as direct photon
production at measured $p_T\gg \Lambda_{\rm QCD}$ the situation is
slightly different.  To leading order in the hard scattering,
incoming partons produce a photon-parton system, which subsequently
evolves into a photon-jet pair. At higher orders in $\as$, the pair
recoils against unobserved soft gluon radiation with total transverse
momentum $Q_T$, in much the same way
as for a single electroweak boson.
When only the photon is observed,
$Q_T$ is integrated and singularities at $Q_T/p_T=0$
cancel, analogously to singularities at $1-z=0$ in threshold
resummation. Thus in the direct photon cross section
at measured $p_T$, both transverse momentum and threshold singularities
are implicit rather than explicit.
Nevertheless, small-$Q_T$
gluon radiation can play a significant role
in the cross section.   Applied to direct photon production, joint
resummation attempts to estimate the effects of these
soft emissions systematically \cite{lae00b,lae00a}.

For an observed photon of momentum $p_T$, the photon transverse
momentum in the pair center-of-mass is
\be
{\bf p}_T' = {\bf p}_T-{\bf Q}_T/2\, .
\label{ptprimedef}
\ee
In the limit that $Q_T / p_T \ll 1$ the cross section
is a convolution \cite{lae00a} of
the resummed distribution in ${Q}_T$ with a hard-scattering
function evaluated at photon momentum $p_T'$.
Non-zero pair momentum $Q_T$, if in the direction
of the observed photon, decreases the scale of the hard scattering,
and can thus enhance the cross section.
As emphasized in Ref.\ \cite{lae00a}, however,
when $Q_T$ grows to the order of $p_T$,
this approximation generates kinematic singularities.
Their effect is non-negligible because
the fall-off in soft gluon transverse momenta has a power-like
perturbative tail. In \cite{lae00a}, we dealt with the
kinematic singularity in a rather crude way by cutting off the resummed
$Q_T$ spectrum at a convenient scale $Q_T=\bar{\mu}$:
\be
p_T^3{ d \sigma_{AB\to \gamma X}^{\rm res} \over dp_T}
=
\int dQ^2 d^2Q_T\; p_T^3
{d \sigma_{AB\to \gamma X}^{\rm res} \over
dQ^2\,
d^2Q_T\, dp_T}\; \theta\left(\bar{\mu}-Q_T\right)\, ,
\label{formal2}
\ee
where $Q$ is the invariant mass of the photon-parton pair. At
threshold, the latter is fixed by
\be
Q= 2p_T \; . \label{Qdef}
\ee
The scale $\bar{\mu}$ in Eq.~(\ref{formal2})
may be regarded as a matching scale.
Ideally, at $Q_T\sim \bar{\mu} < p_T$, one would replace the resummed cross
section $p_T^3 {d \sigma_{AB\to \gamma X}^{\rm res}/dQ^2\,
d^2Q_T\, dp_T}$ by the fixed-order (NLO) one,
which does not have the kinematic singularity. In practice,
this becomes a very complicated procedure, and it is more
convenient to derive a jointly resummed cross section that does
not require a cutoff. We will show below that this may be achieved
by applying an additional, self-consistent approximation
that is exact at partonic
threshold.  To do so, we must recall the explicit form of the cross
section derived in Refs.\ \cite{lae00b,lae00a}.

\subsection{The double inverse transform}

Integrated over rapidities, the jointly resummed direct photon
cross section is written in terms of Mellin moments of the
$\overline{\rm MS}$ parton distributions, $\tilde \phi_{a/H}(N,\mu)
\equiv \int_0^1 dx x^{N-1}\, \phi_{a/H}(x,\mu)$, as
\bea
p_T^3 \, {d \sigma^{({\rm resum})}_{AB\to \gamma X} \over dp_T}
&=& \sum_{ab} \frac{p_T^4}{8 \pi S^2} \int_{\cal C} {dN \over 2 \pi i}\;
\tilde{\phi}_{a/A}(N,\mu) \tilde{\phi}_{b/B}(N,\mu)\;
\int_0^1 d\tilde x^2_T \left(\tilde x^2_T \right)^N
{|M_{ab}(\tilde x^2_T)|^2\over \sqrt{1-\tilde{x}_T^2}}
\nonumber \\
&& \hspace{-15mm} \times \ C^{ab\to \gamma c}(\as(\mu),\tilde{x}_T^2)
\int {d^2  Q_T \over (2\pi)^2}\;
\Theta\left(\bar{\mu}-Q_T\right)
\left( \frac{S}{4  p_T'{}^2} \right)^{N+1}\;
P_{ab}\left( N,Q_T,\frac{2 p_T}{\tilde x_T},\mu \right)\, ,
\label{1pIresumE}
\eea
where $\mu$ is the factorization and renormalization scale,
and the $|M_{ab}|^2$ are squared amplitudes for the partonic
processes $ab\to \gamma c$.
The variable $\tilde x_T^2$ is defined by
\bea
\tilde x_T^2 \equiv \frac{1}{\cosh^2\tilde\eta}\, ,
\eea
where $\tilde\eta$ is the rapidity of the direct photon in
the center of mass of the hard scattering.  At partonic
threshold, or equivalently large values of the moment
variable $N$, $\tilde\eta$ is forced to unity.
For this reason, we will approximate
\bea
2p_T/\tilde x_T\sim 2p_T\equiv Q
\eea
in the functions $P_{ab}$ in Eq.\ (\ref{1pIresumE}), where
dependence on $p_T$ is logarithmic.
The contour ${\cal C}$  in Eq.\ (\ref{1pIresumE}) and the
$b$ integral in (\ref{Pdef}) below define the inverse transforms
from $N,b$ space to $z$ and $Q_T$. These contour integrals
were described in detail in Refs.\ \cite{lae00a,kul02}.

The functions $P_{ab}$ in Eq.\ (\ref{1pIresumE})
were derived in Ref.\ \cite{lae00a} and provide $Q_T$
dependence at fixed $N$. Each $P_{ab}$ is itself the Fourier
transform of the exponentiated logarithmic dependence on $N$ and $b$,
\bea
P_{ab}\left( N,Q_T,Q,\mu \right)
=
\int d^2  b \,
{\rm e}^{-i  b \cdot Q_T} \,
\exp\left[E_{ab\to \gamma c}\left( N,b,Q,\mu \right)\right]\, ,
\label{Pdef}
\eea
where the $E_{ab\to \gamma c}$ are ``Sudakov" exponents that
we will specify explicitly below.
They can be split into initial and final state
contributions, where, as shown in \cite{lae00a},
all $b$-dependence comes from the initial state,
\be
E_{ab\to \gamma c}(N,b,Q,\mu)
=E_{ab}^{\rm IS}(N,b,Q,\mu) + E_{abc}^{\rm FS}(N,Q,\mu)\, .
\label{splitE}
\ee
The $N$-independent coefficients $C^{ab\to \gamma c}$ contain the
effects
of hard virtual corrections and are perturbative series
of the form $C^{ab\to \gamma c}=1+\frac{\alpha_s}{\pi}
C^{ab\to \gamma c\, (1)}+\ldots$. To next-to-leading
logarithmic accuracy one needs the first order terms which
may be found in \cite{cmnov}, and are given in Appendix\ A below.
In this paper, we concentrate on the initial state exponent,
which contains all leading logarithmic effects and all
$b$ dependence.

Threshold resummation is recovered from Eq.\ (\ref{1pIresumE})
by setting $b$ to zero
in the exponents $E_{ab\rightarrow
\gamma c}$.
In this case, the $b$ integral produces a delta function that
sets $p_T'=p_T$.  Then the exponent $E_{ab\rightarrow \gamma c}$
reverts to its threshold resummed form, and $S/4p_T'{}^2 \rightarrow
1/x_T^2$. Recoil enhances the cross section
Eq.\ (\ref{1pIresumE}) because even for $Q_T\ll p_T$, the
ratio $S/4p_T'{}^2$ can be larger than $S/4p_T^2$.
For large enough $Q_T$, the factor
$S/4p_T'{}^2$ can diverge, and a cutoff
is required, as discussed above.  This momentum
configuration, however, requires, $Q_T \sim 2p_T$,
and hence is far outside the region where resummation
is applicable. This problem is not due to
our approximation in the region of interest, but
to our extrapolation beyond that region.
When $Q_T$ approaches $p_T$ in magnitude, the factorization
between gluon emission and hard-scattering
fails. For large $N$, however,  the ``profile'' functions $P_{ab}$
vanish once $Q_T>Q/N \ll p_T$. The $P_{ab}$'s vanish for moderate
$Q_T$  because the exponents $E_{ab\rightarrow \gamma c}$ develop
large (negative) logarithms once $bQ/N>1$ (see
below, Eq.\ (\ref{fullexp})).  This ensures
that for $Q_T>Q/N$ the exponential $\exp[-ib\cdot Q_T]$ oscillates
on a smaller scale than the size of the region where
$\exp[E_{ab\rightarrow \gamma c}]$ is nonvanishing.
Put another way, because the widths of the profile functions in $b$ space
are of order $N/Q$, their Fourier transforms to $Q_T$ space
have widths of order $Q/N$.
Numerical examples for the $Q_T$ integrand in Eq.~(\ref{formal2})
were given in Ref.\ \cite{lae00a}, which show the fall-off
of the profile function for increasing $Q_T$, followed
by the kinematic singularity as $Q_T$ increases
to the order of $p_T$. We conclude that for large $N$ the true
enhancement due to
recoil is insensitive to modifications of the integrand above $Q_T \sim
p_T/N \ll p_T$.
Since large $N$ corresponds to $x_T\rightarrow 1$, we expect a
suppression of recoil effects in this limit.  In addition, $N$ is
conjugate to $k_0/p_T$, where $k_0$ is the energy of initial
state radiation \cite{lae00a,lae00b}.  The relation $Q_T<p_T/N$ is thus
equivalent to the restriction that the total transverse momentum of
initial state radiation is less than its energy.
We will use this observation shortly.

\subsection{Elimination of the kinematic singularity}

Given that all-order recoil effects enhance the jointly
resummed cross section from values of $Q_T$
such that $Q_T < p_T/N$, it is only in this region that
we are required to maintain accurate expressions for
leading $Q_T$ behavior (that is, $1/Q_T^2$ times logarithms).
In fact, to construct the jointly resummed
expression in Eq.\ (\ref{1pIresumE}), we have neglected
corrections that are nonsingular at $Q_T=0$
and $1-z=0$.  This means that we do not
in general have control over corrections
suppressed by powers of $Q_T/p_T$, and also that
we are free to change the resummed expression at
this level of accuracy.  Such a change will only affect
the result from the region of $Q_T$ beyond the
range that gives enhancement.   These modifications will
not produce logarithms, and
we can adjust for them by matching to the cross section at
fixed order.

In summary, we are free to
choose an extrapolation that does not produce spurious
singularities at large $Q_T$ and which does not change the
singularity structure at $Q_T=0$.  This may be done in such a
way that the resulting resummed expression remains accurate
to NLL in the variables $N$ and $b$.

In this spirit, we make the following
approximation, accurate to corrections that
are suppressed by
factors of $Q_T/p_T$:
\bea
\left( \frac{S}{4 (\vec{p}_T-\frac{1}{2}\vec{Q}_T)^2}\right)^{N+1}
=
\left(x_T^2\right)^{-N-1}\; \exp\; \left\{ (N+1) \, \vec{Q}_T\cdot
\vec{p}_T/p_T^2
\ \left[1+{\cal O} \left(\frac{NQ^2_T}{p^2_T}\right) \, \right]\,
\right\}\, .
\label{replace1}
\eea
Notice that the exponent reaches order unity at just
those values of $Q_T$ for which the profile function
begins to decrease.  Replacing the singular power
dependence on $Q_T$ with the exponential, we retain
the leading behavior at low $Q_T$, but eliminate the
kinematic singularity, as desired. Again we emphasize that
suppression for $Q_T>p_T/N$ is a reflection of energy conservation.

Let us now study the effect of the approximation in Eq.\
(\ref{replace1}).
Consider for the moment $N+1=-i{\cal N}$ with ${\cal N}$ fixed and real.
Using Eq.~(\ref{Pdef}), we may then replace
the second line of Eq.~(\ref{1pIresumE}) according to
\bea
C^{ab\to \gamma c}\,\int {d^2 {Q}_T \over (2\pi)^2}\;
\Theta\left(\bar{\mu}-Q_T\right)
\left( \frac{S}{4 {p}_T'{}^2} \right)^{N+1}\;
P_{ab}\left( N,Q_T,\frac{2 p_T}{\tilde x_T},\mu \right)
&\ &
\nonumber \\
&  & \hspace{-90mm} \longrightarrow \,
C^{ab\to \gamma c}\,\left(x_T^2\right)^{-N-1}\;
\int d^2 {b} \,
\exp\left[E_{ab\to \gamma c}\left( N,b,Q,\mu \right)\right]
\int {d^2 {Q}_T \over (2\pi)^2}\;
{\rm e}^{-i{\cal N} \vec{Q}_T\cdot \vec{p}_T/p_T^2-i \vec{b}\cdot
\vec{Q}_T}\, .
\nonumber\\
\label{replace2}
\eea
Here we have extended the $Q_T$ integral to infinity. The
integral may then be performed, and gives $\delta^{(2)}
\left( \vec{b} + {\cal N} \vec{p}_T/p_T^2 \right)$.
Using this delta function to perform the $b$ integral
in (\ref{replace2}), and inserting the result
back into Eq.~(\ref{1pIresumE}), we find
\bea
p_T^3 \, {d \sigma^{({\rm resum})}_{AB\to \gamma X} \over dp_T}
&=& \sum_{ab} \frac{p_T^4}{8 \pi S^2} \int_{\cal C} {dN \over 2 \pi i}\;
\tilde{\phi}_{a/A}(N,\mu) \tilde{\phi}_{b/B}(N,\mu)\;
\int_0^1 d\tilde x^2_T \left(\tilde x^2_T \right)^N
{|M_{ab}(\tilde x^2_T)|^2\over \sqrt{1-\tilde{x}_T^2}}
\nonumber \\
&& \hspace{-5mm} \times \,
C^{ab\to \gamma c}\,\left(x_T^2\right)^{-N-1}\;
\exp\left[E_{ab\to \gamma c}\left( N,-i\frac{N+1}{ p_T},Q,\mu
\right)\right] \; .
\label{profilefinal}
\eea
This expression for the direct photon cross section
is similar to the result for pure threshold
resummation, except for the additional $b$ dependence, which
has become dependence on the combination $(N+1)/p_T$ in the exponent.
Although we have derived this form for imaginary values of $N+1$, it
can be analytically continued to any $N$,
and we use (\ref{profilefinal}) as the result of the $Q_T$
integral in the high-$x_T$ jointly resummed cross section.
Recoil is self-consistently taken into account through
the exponential in (\ref{replace1}), which is accurate up
to power corrections as shown.  There are no kinematic
singularities at large $Q_T$.
For $Q_T$ competitive with $p_T$, of course, the approximations we have
made fail, but in this region the profile function is small.

\subsection{Matching}

Matching is now straightforward for the stabilized cross section, Eq.\
(\ref{profilefinal}),
and can be handled as for the threshold-resummed cross section.
We simply expand the exponents to NLO (for example) in
terms of $\alpha_s(p_T)$, and replace these approximate
expressions with the exact hard scattering cross sections at
that order.

We emphasize that within our new treatment we have
been able to perform both the $Q_T$ and the $b$ integrals.
This  is a great
advantage for phenomenological applications, since now
the evaluation of the cross section is technically equivalent
to that of a standard threshold-resummed one. In fact,
as we show below, Eq.\ (\ref{profilefinal}) is identical
to normal threshold resummation to NLL, but
differs at NNLL through a well-defined set of
terms that can be identified uniquely as recoil effects.
The fact that the NLL threshold logarithms are unchanged
by recoil is an important consistency check of our approach
because these logarithms are uniquely specified in the perturbative
single-inclusive cross section.

The final resummed cross section thus has a form that is closely
related to
matched threshold resummation \cite{kul02}:
\bea
p_T^3\,\frac{d\sigma_{AB}^{\mathrm{res}}}{dp_T}
&=&\sum_{ab}
\int_{\cal C} \, \frac{dN}{2\pi i} \;
\tilde{\sigma}_{ab}^{(0)}(N)\;C^{ab\to \gamma c}\;
\left(x_T^2\right)^{-N-1} {\rm
e}^{E_{ab\to \gamma c}\left( N,-i\frac{N+1}{ p_T},Q,\mu
\right)}\nonumber \\
& & + \; p_T^3\,\frac{d\sigma_{AB}}{dp_T}^{\mathrm{NLO}}\;-\;
p_T^3\,\frac{d\sigma_{AB}}{dp_T}^{\mathrm{res}|_{\alpha_s^2}} \; ,
\label{newsigma}
\eea
where $\sigma^{(0)}_{ab}(N)$ is the moment of
the lowest-order cross section,
\bea
\tilde{\sigma}_{ab}^{(0)}(N)\;=
\frac{p_T^4}{8 \pi S^2}\;
\tilde{\phi}_{a/A}(N,\mu) \tilde{\phi}_{b/B}(N,\mu)\;
\int_0^1 d\tilde x^2_T \left(\tilde x^2_T \right)^N
{|M_{ab}(\tilde x^2_T)|^2\over \sqrt{1-\tilde{x}_T^2}}\, ,
\eea
and where the final terms in (\ref{newsigma})
express our matching to the fixed
order (NLO, ${\cal O}(\alpha_s^2)$) cross section
$p_T^3\,\frac{d\sigma_{AB}}{dp_T}^{\mathrm{NLO}}$ by
taking out the ${\cal O}(\alpha_s^2)$ expansion of the
perturbative part of the resummed cross section,
$p_T^3\,\frac{d\sigma_{AB}}{dp_T}^{\mathrm{res}|_{\alpha_s^2}}$.

\section{Perturbative and Nonperturbative Exponents}

\subsection{Resummed perturbative recoil}

To clarify the relationship between joint and threshold resummation
and the implications of our new treatment of recoil, we review the
$N$ and $b$ dependence of the resummed exponent at NLL found in
\cite{lae00b}.
To all orders, the NLL initial-state logarithms in $N$ and $b$ are
generated
from an integral, derived using the eikonal nature of
soft gluon emission, that extends down
to zero scale in the running coupling.  It may be written
in a convenient form as
\bea
E_{ab}^{\rm IS}(N,b,Q,\mu=Q)  &\ & \label{Eelaborate} \\
&\ & \hspace{-20mm}  =
\int_0^{Q^2} {d k_T^2\over k_T^2}\;
\sum_{i=a,b} A_i\left(\as(k_T)\right)\;
\left[\, J_0 \left( b k_T \right) \;
K_0\left({2Nk_T\over Q} \right) + \ln\left({\bar N k_T\over
Q}\right)\, \right]\,\nonumber
\, ,
\eea
where $Q\equiv 2p_T$ (see Eq.~(\ref{Qdef})) is the minimal center of
mass
energy of the partonic subprocess. Here and below, we define
\bea
\bar{N}=N {\mathrm{e}}^{\gamma_E}\, .
\eea
The anomalous dimensions $A_a(\as)$ have the familiar
expansion $A_a(\as)=\sum_n\, (\as/\pi)^n\; A_a^{(n)}$, with
\bea
A_a^{(1)} &=& C_a \nonumber \\
A_a^{(2)} &=& \frac{1}{2} C_a K \equiv \frac{1}{2} C_a \left[ C_A \left(
\frac{67}{18} - \frac{\pi^2}{6} \right) -\frac{10}{9} T_R N_f \right]\,
,
\label{explicitA}
\eea
where $C_q=C_F$ for quarks and $C_g=C_A$ for gluons.
The presence of the Bessel function $K_0(2Nk_T/Q)$ reflects
the conservation of energy that must be imposed to
resum threshold and $k_T$ enhancements simultaneously.
For the analogous exponent in $k_T$ resummation,
the function $K_0(2Nk_T/Q)$ in Eq.\
(\ref{Eelaborate}) is replaced by $-\ln(k_T/Q)$
and $\ln({\bar{N}} k_T/Q)$ by $\ln(k_T/Q)$, and the
$k_T$ integral produces
logarithms of $bQ$ for any $b>1/Q$.  In joint resummation,
however, $b$ must be greater than $N/Q$ to produce logarithms.
As a result, for $N\rightarrow\infty$ the profile function in $Q_T$
space
decreases once $Q_T > Q/N$, as discussed in Sec.\ 2.2 above.

Starting from Eq.\ (\ref{Eelaborate}) we isolate the effect of
perturbative recoil by separating it from the corresponding
exponent for threshold resummation.  Since threshold resummation
is already accurate to NLL in the transform variable $N$ \cite{lae00b},
for consistency recoil must appear first at the next logarithmic order,
and it does. As we shall see, however, its influence on
the 1PI cross section need not be negligible
in perturbation theory.  In addition, the integral over
the anomalous dimension $A(\as(k_T))$ through the infrared region
suggests a specific set of nonperturbative corrections,
whose effects we will also study. For initial-state radiation,
the form of contributions beyond NLL accuracy is given in \cite{lae00b}.
Threshold logarithms associated with final-state interactions
beyond NLL will be the subject of a separate investigation. We will
argue that they respect the pattern for power corrections found here.

As directed by Eq.~(\ref{profilefinal}), we now
set $b=-i(N+1) / p_T$ in Eq.~(\ref{Eelaborate}),
noting the Bessel function relation $J_0(iz)=I_0(z)$. We
then reorganize the equation as
\bea
E_{ab}^{\rm IS}(N,b=-i\frac{N+1}{p_T},Q,\mu=Q)  &&  \nonumber \\
&\ &\hspace{-40mm}=
\int_0^{4p_T^2} {d k_T^2\over k_T^2}\;
\sum_{i=a,b} A_i\left(\as(k_T)\right)\;
\left[\, K_0\left({Nk_T\over p_T} \right) +
\ln\left({\bar N k_T\over {2p_T}}\right)\, \right]\nonumber\\
&\ &\hspace{-35mm}+
\int_0^{4p_T^2}\frac{dk_T^2}{k_T^2}\;
\sum_{i=a,b} A_i\left(\as(k_T)\right)\;
\left[ I_0\left(\frac{(N+1)k_T}{p_T}\right)-1\right]
K_0\left(\frac{N k_T}{p_T}\right) \,\nonumber\\
&\ &\hspace{-40mm}\equiv E_{ab,\, \mathrm{thr}}^{\rm IS}(N,p_T)+
\,\delta\,E_{ab,\, \mathrm{rec}}(N,p_T) \, ,
\label{fullexp}
\eea
where  we have again used $Q=2p_T$.
We have identified the first term on the right-hand-side
of Eq.~(\ref{fullexp}) with the exponent for threshold
resummation for initial-state logarithmic behavior in
$N$~\cite{lae00b}.

The second term on the right side of
(\ref{fullexp}) is the recoil correction. It now amounts
simply to an $N$-dependent correction to the threshold-resummed
cross section. As required by the self-consistency of NLL
threshold resummation, this expression is free of NLL logarithms
in $N$, because for small arguments $z$,
\bea
I_0(z) \sim 1 + \frac{z^2}{4}\, , \quad \quad
K_0(z) \sim
     -\ln \left[ \frac{z{\rm e}^{\gamma_E}}{2}\right]\,
     \left( 1+ {z^2\over 4}\right) +{z^2\over 4} \, .
     \label{smallN}
\eea
On the other hand, as $z\equiv N k_T/p_T$ becomes large
with ${\rm Re}(z)>0$, $I_0(z)$
increases as ${\rm e}^z/\sqrt{2\pi z}$, while $K_0(z)$
decreases as ${\rm e}^{-z}/\sqrt{(2z/\pi)}$, so that
\bea
I_0(z)K_0(z) \rightarrow \frac{1}{2z} \quad\quad({\rm Re(z)>0})\ \, .
\label{largeN}
\eea
At fixed coupling, and replacing $N+1$ by $N$ in $I_0$ in
Eq.~(\ref{fullexp}), the net result is a convergent,
$N$-independent integral, equal to $\left( C_a+C_b
\right)$ $\times$ $(\alpha_s/2\pi)$ $\zeta(2)$, a
modest but still significant contribution
in the exponent.

For large values of $N$, we can readily
estimate the effect of the running coupling to
perturbative recoil,  by noting that the
combination $(I_0-1)K_0$ becomes sharply peaked near
$k_T = p_T/N$, with a width that is asymptotically negligible
compared to the scale on which the coupling runs.  As a result,
to NNLL, we may isolate perturbative recoil including the
running of the coupling by the expression,
\be
\delta\,E_{ab,\, \mathrm rec}^{\rm NNLL} = \left( C_a+C_b \right)
{\alpha_s(4p_T^2/\bar{N}^2) \over
2\pi}\ \zeta(2)\; .
\label{zeta2}
\ee
We will use this expression below to estimate the effects
of perturbative recoil.

\subsection{Nonperturbative corrections from threshold}

The recoil exponent $\delta\,E_{ab,\,  \mathrm{rec}}(N,p_T)$ in
Eq.~(\ref{fullexp}) provides an estimate of
perturbative recoil, and is also a guide to nonperturbative power
corrections. The most basic observation about these corrections
is that they factorize and exponentiate,
in much the same manner as for event shapes in $\rm e^+e^-$
annihilation and for the transverse momentum distributions
of electroweak boson production.   This follows from the form
of the resummed exponent, in which the entire dependence on
the running coupling is through a  single, integrated scale, $k_T$.
We emphasize that a similar result holds for the full eikonal
exponent to all logarithmic order.  Indeed, the same underlying
nonperturbative parameters that appear in
Drell-Yan cross sections will appear in power corrections
to direct photon cross sections.
As noted in \cite{kul02}, power corrections
from threshold and transverse momentum resummations
are separately additive in the exponent.   As we shall see,
this leads to an extra power correction compared
to estimates based on transverse momentum resummations
alone \cite{ktresum1,ktresum2}.

Using the additivity of the nonperturbative corrections, we write
for the full exponent
\bea
E_{ab\to \gamma c}(N,p_T) &=& E_{abc}^{\rm  PT} +
\delta\, E_{ab}^{\mathrm np} \; ,
\nonumber\\
E_{abc}^{\rm  PT} &=&  E_{abc,{\rm thr}}^{\rm PT} +
\delta\,E_{ab,\, \mathrm rec}^{\rm NNLL}\; ,
\eea
where $\delta\, E_{ab}^{\mathrm  np}$ accounts for nonperturbative
contributions from low scales in $k_T$, of order $\Lambda_{\rm QCD}$.
The full  perturbative threshold exponent at NLL, $E_{abc,{\rm
thr}}^{\rm PT}$, with initial- and final-state
contributions, was derived in \cite{cmnov,kido,sv,cat96}.
As noted above, in this study we derive nonperturbative
and recoil corrections associated with initial-state
radiation only.

For small to moderate values of $N$, the integral
in $E^{\mathrm IS}_{ab}(N,p_T)$ is perturbatively dominated.
Nonperturbative corrections are generated by treating
$Nk_T/p_T$ as a small parameter in both the
threshold and recoil exponents of Eq.\ (\ref{fullexp}).  Expanding the
integrands of both $\delta\,E_{ab,\, {\mathrm{rec}}}(N,p_T)$
and $E^{\mathrm IS}_{ab, \,\mathrm  thr}$ in Eq.\ (\ref{fullexp})
for small $k_T^2$, we parameterize the resulting $1/p_T^2$ terms as
\bea
\delta\,E_{ab}^{\mathrm{np}}&=&\frac{(N+1)^2+N^2}{4p_T^2}\sum_{i=a,b}
\left[\lambda^{1,1}_i + \lambda^{1,0}_i \ln
\left(\frac{2p_T}{\kappa\bar{N}}\right)\right] + \frac{N^2}{4p_T^2}
\sum_{i=a,b}\, \lambda_i^{1,0}\quad \;
(N\kappa/p_T < 1)\, ,
\label{powernp}
\eea
where the $N^2$ terms come from the threshold ($K_0$) integral
in Eq.\ (\ref{fullexp}), while the $(N+1)^2$ term is from
the recoil ($I_0$) term.  The logarithm in both
cases arises from the expansion of the function $K_0(2Nk_T/p_T)$,
and, as noted above, its presence can be traced to
the imposition of energy conservation in joint resummation.
The constants $\lambda_i^{m,n}$ in Eq.\ (\ref{powernp})  are
interpreted as
the nonperturbative content of moments of the
running coupling \cite{disper95,disper99,kor95}, with indices in
a  notation inspired by \cite{disper99}.
More specifically, these are moments of  the anomalous
dimensions $A_a\left(\alpha_s(k_T)\right)$, \cite{kor99}
\bea
\lambda_a^{m,n} &=& \int_0^{\kappa^2} dk_T^2 \left(k_T^2\right)^{m-1}\;
A_a\left(\alpha_s(k_T)\right)\,
\ln^n\left(\frac{k_T}{\kappa}\right)\nonumber\\
&=& \frac{C_a}{\pi} \;
\int_0^{\kappa^2} dk_T^2 \left(k_T^2\right)^{m-1}\;
\alpha_s(k_T)\, \ln^n\left(\frac{k_T}{\kappa}\right)+ \dots\, .
\label{npmom}
\eea
In Eq.\ (\ref{npmom}), the upper
limit $\kappa$ for the $k_T^2$ integral, which also appears as the scale
in logarithms of $p_T$ in Eq.\ (\ref{powernp}),
is a factorization scale.
To isolate a truly nonperturbative coupling, as in Ref.\
\cite{disper99},
we could subtract perturbative contributions to the $\lambda$'s
to the order corresponding to our level of resummation.
Since this process does not change the $p_T$ and $N$ dependence
of the expressions, and because the nonperturbative
parameters appear in the same manner here as in
electroweak annihilation \cite{ktresum1}, it is not
necessary to provide such an analysis
for our purposes.

As in the case of electroweak bosons \cite{dynp,bro02,dyeven}, and in
contrast to
event shapes \cite{eventpc,disper95,kor95,shape},
only even powers result from the expansion of the Bessel functions in
(\ref{fullexp}), the first of which has been displayed in
Eq.~(\ref{powernp}).
For $N$ not too large, that is, for
$N\Lambda_{\rm QCD}\ll p_T$ we expect only
one or two power corrections to be significant, but for larger $N$,
the resummed cross section should be supplemented by
a function with a more general $N$-dependence \cite{kor99,shape}.
In the following section, we will study
the phenomenological implications of such dependence.

\subsection{The full exponent}

Summarizing our results so far,
the full exponent is the sum of a perturbative threshold exponent,
perturbative recoil
and nonperturbative corrections,
\bea
E_{ab\to \gamma c}\left( N,i\frac{N+1}{ p_T},Q,\mu\right)
&=&
E_{abc, \mathrm{thr}}^{\rm  PT}(N,p_T) + \delta\, E_{ab,\,
\mathrm{rec}}^{\mathrm{NNLL}}(N,p_T)
+ \delta\,E_{ab}^{{\mathrm np}} \, ,
\label{fullexpall}
\eea
where the nonperturbative exponent
$\delta E_{ab}^{\mathrm np}$ is given in (\ref{powernp}) above and
the NNLL recoil correction $\delta\,E_{ab,\,
\mathrm{rec}}^{\mathrm{NNLL}}$ by (\ref{zeta2}).
$E_{ab,\, \mathrm{thr}}(N,p_T)$ is the full exponent for
threshold resummation in prompt-photon production,
including initial and final state contributions
\cite{cmnov,kido,sv,cat96} (see Appendix\ A).

\section{Phenomenology of Power Corrections}

The expressions derived above provide useful
information on the phenomenology of power corrections
associated with soft gluon emission.  First,
$p_T$ dependence enters through
even powers, with a leading nonperturbative coefficient that
is identical to that encountered as the
coefficient of $b^2$ in electroweak
annihilation.   For comparison, the latter
may be written in terms of
the same parameters $\lambda_q^{m,n}$ as
\bea
\delta\,E_{q\bar{q}}^{\mathrm{np}}\ \mathrm{(Drell-Yan)} &=& -{b^2\over
4}\;
\sum_{i=q,\bar{q}}
\left[\lambda^{1,1}_i + \lambda^{1,0}_i \ln
\left(\frac{Q}{{\kappa}}\right)\right]\; .
\label{powernpdy}
\eea
In contrast, for the single-particle inclusive cross section
in joint resummation
the nonperturbative corrections
in Eq.\ (\ref{powernp}) possess highly nontrivial $N$-dependence,
from recoil directly, as well as from threshold
resummation.    This implies that these power corrections
inherit nontrivial energy dependence, and we may
expect their effects to change with the overall
energy.

To see the qualitative energy-dependence
implied by the nonperturbative exponents derived
above, we note that  the
Mellin moment $N$ and the variable $\ln x_T^2$ are in a conjugate
relationship, exhibited in the inverse transform (\ref{newsigma}),
\be
N \; \leftrightarrow \;\frac{1}{\ln x_T^2}\; .
\label{conjugate}
\ee
Identifying these quantities in the nonperturbative
power correction of Eq.\ (\ref{powernp}),
and recalling that $x_T =2p_T/\sqrt{S}$, we immediately see
that the nonperturbative exponent is suppressed not only
by a power of $p_T$, but also by a power of $\ln S$ at fixed $p_T$,
\be
\delta\, E_{ab}^{\rm np}\; \leftrightarrow \;
\frac{\lambda^{1,0}_a+\lambda^{1,0}_b}{4p_T^2\ln^2\left(\frac{4p_T^2}{S}
\right)}
\;\ln\left[ 2p_T\ln\left(\frac{4p_T^2}{S}\right)\right]\, .
\label{lnssuppress}
\ee
Even though this expression is eventually to be
convoluted with the hard scattering cross sections
and the parton distributions,
we may conclude that at fixed
$p_T$, the importance of power corrections will decrease
as $\sqrt{S}$ increases. At the same time, as $x_T$ approaches
unity, the coefficient of $1/p_T^2$ diverges, and the
nominal power correction may dominate at the edge of
phase space. Even as this coefficient diverges, however,
the logarithm in the numerator eventually changes sign, so that for
$x_T$ close enough to unity the enhancement becomes a
suppression.  This is not an accident, because the
presence of the factor of $N$ in the logarithm reflects
energy conservation, which is respected by joint resummation.
To give a realistic estimate of
the behavior of these nonperturbative corrections,
we return to moment space.

For the dominant form of the
nonperturbative exponent at moderate
$N$ and $p_T$, we are guided by Eq.~(\ref{powernp}).
For larger $N$, however, all power corrections in $N$ may
become relevant. To account
for this, we introduce a function of $N/p_T$ that
generalizes this expression.  We will refer
to this as a shape function by analogy to the
discussion of \cite{kor99}. We
tailor the $N$-dependence to the  behavior
of the Bessel functions of Eq.\ (\ref{fullexp})
for large and small values of their arguments,
which  both depend on the combination $N/p_T$.
Matching to the small-$N$ behavior
of the Bessel functions in (\ref{smallN}) and
to the large-$N$ behavior in (\ref{largeN}), we modify Eq.\
(\ref{powernp}),
\bea
\delta E_{ab}^{\mathrm np} =  \mu_0\,
\frac{C_{a}+C_{b}}{\pi}\;{(N+1)^2+ N^2 \over 4p_T^2}\;
{ \ln\left (1+{2p_T \over \bar NQ_0}\right) \over \left( 1+ {Q_0N\over
p_T}
\right)^2}\, .
\label{shapenp}
\eea
Here the scale $\mu_0$ is a parameter of dimension
mass squared, which can be thought of as the
integral of $A_a(\as(k_T))(\pi/C_a)$ over $k_T^2$
with unit weight in (\ref{npmom}).
The overall factors of color charges in
Eq.\ (\ref{shapenp}) reflect the proportionality of the coefficients
$A_a$ to $C_a$.
The parameter $Q_0$ is a scale whose value accounts for non-logarithmic
terms.
If
$\lambda^{1,1}=0$
in (\ref{npmom}),
we may identify $Q_0$ with
the scale $\kappa$ in Eq.\ (\ref{npmom}).  This is the
result found in Ref.\ \cite{disper99} with $\kappa=2$ GeV.

To estimate the impact of these nonperturbative corrections
for large $x_T$, we are aided by our experience with electroweak boson
production and with perturbative resummation. As can be
seen from Eq.~(\ref{npmom}), the nonperturbative parameters
in our approach are related to moments of the strong coupling,
which suggests some form of universality for them.
In a study of Z production at the Tevatron we estimated
nonperturbative effects; the value obtained may be translated
into [($\lambda^{1,0}_{q}+\lambda^{1,0}_{\bar{q}})/4]\,
\ln(m_{\mathrm{Z}}/\kappa)\approx 0.8$
GeV$^2$ in Eq.\ (\ref{shapenp}),
or $\lambda^{1,0}_q\sim \lambda^{1,0}_{\bar{q}}\sim 0.4$~GeV$^2$,  
consistent
with the result quoted in \cite{disper99} for $\kappa=2$ GeV.
This implies for the parameter $\mu_0$ in (\ref{shapenp}),
\bea
\mu_0\ {\mathrm{(Drell-Yan)}} = {\pi \lambda^{1,0}_q\over C_F}\, \sim 1
\
{\mathrm{GeV}}^2 \; .
\eea
The nonperturbative exponent $\delta E_{ab}^{\mathrm np}$
in Eq.\ (\ref{shapenp}) with this value of $\mu_0$
is our best estimate of the $N$-dependent exponent at
large $N$.  Substituted into the full exponent (\ref{fullexpall})
it provides a measure of power corrections at large $x_T$.

To illustrate the influence of these power corrections, we
compute the ratio of the cross section with
threshold resummation plus the nonperturbative term (\ref{shapenp})
to the cross section with threshold resummation alone. We do this
for several cases that are directly relevant for comparison with
experiments: for $p\,Be$ scattering with fixed-target beam energies
$E=530$~GeV and $E=800$~GeV (E706 \cite{e706data}),
for $pp$ and $\bar{p}p$ scattering with beam energy $E=315$~GeV
(UA6 \cite{ua6data}), and for $pp$ scattering at $\sqrt{s}=63$~GeV
(R806 \cite{r806data}).
We use the GRV set of parton distributions from Ref.\ \cite{grv}. We  
normalize
all our results to the threshold resummed cross section,
i.e., Eq.~(\ref{newsigma}) with $\delta\,
E_{\mathrm{recoil}}^{\mathrm{NNLL}}=\delta\,E_{\mathrm np}=0$.
This is advantageous because the dependence on the factorization
and renormalization scales is small in this cross section, and
because our recoil and nonperturbative
corrections have been defined relatively to it.
Note that as implied by Eq.~(\ref{1pIresumE}) our resummation is
done for the cross section integrated over all photon
rapidities. In principle, we should account for the finite
ranges of rapidity covered in the various experiments, which could
be done using the techniques developed in Ref.~\cite{sv}. However,
as implied by the results of~\cite{sv},
the dependence on rapidity will be very
weak in the ratios we consider here and can be neglected for
simplicity. Our results are always matched to the NLO cross section
as described after Eq.~(\ref{newsigma}). We do not take into
account a photon fragmentation contribution to the cross section.

\begin{figure}[h]
\centerline{\epsfig{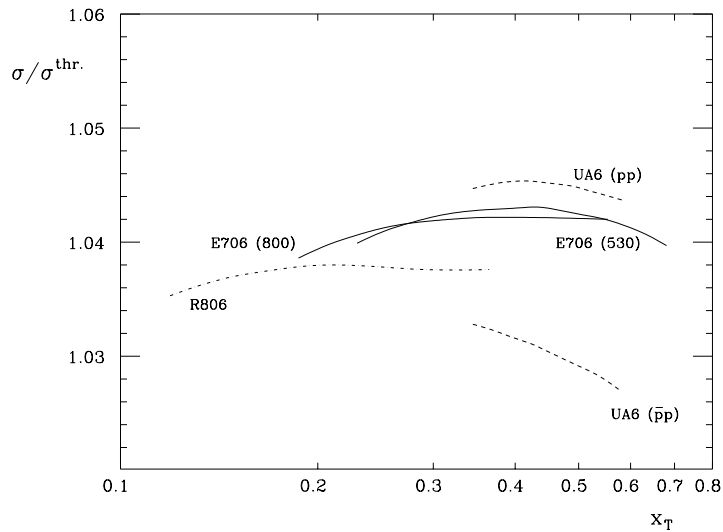}}
\caption{Ratios of direct photon cross sections
computed with threshold resummation and nonperturbative shape function
(\ref{shapenp}) to cross sections with threshold resummation only.
The curves are given as functions of $x_T$ for kinematics
relevant for comparison to fixed-target and ISR experiments (see
text).
\label{ratioshape}}
\end{figure}

Shown in Fig.\ \ref{ratioshape} are results
for the energies discussed above, as functions of $x_T$.
The enhancements exhibited in the figure are both small and
for the most part relatively flat. We have
included values of $x_T$ far from unity, especially for R806,
to illustrate the point that these power
corrections decrease with energy.  For RHIC and CDF
energies, the effects of (\ref{shapenp}) are practically negligible,
of the order of just a few tenths of a percent.  We emphasize, however,
that this result applies only to extrapolations to
small $x_T$ of the expressions derived for $x_T\rightarrow 1$,
and is not necessarily representative of the true
behavior of the cross section at low $x_T$.

\begin{figure}[h]
\centerline{\epsfig{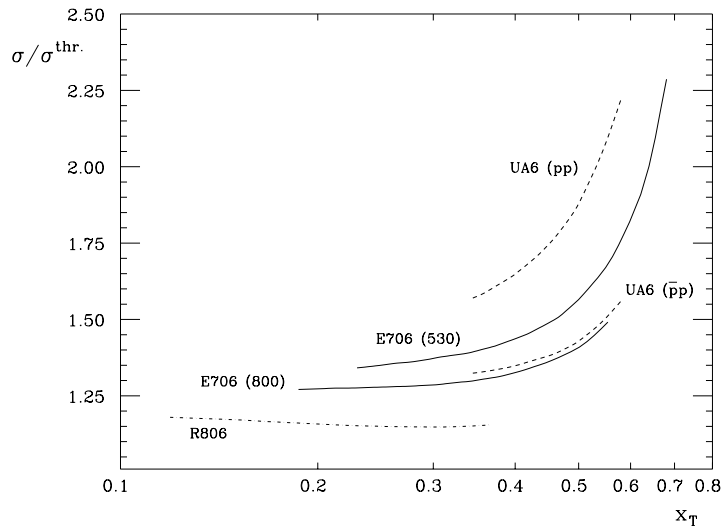}}
\caption{Same as Fig.~\ref{ratioshape}, but for the
$k_T$-inspired nonperturbative shape function
(\ref{shapenpbonly}). \label{shapevsn2}}
\end{figure}

The moderation of the $x_T$ dependence of the cross
section at large $x_T$
in Fig.\ \ref{ratioshape} associated with energy conservation
is illustrated by comparison to
Fig.\ \ref{shapevsn2}, which shows the analogous ratios
when the
nonperturbative coefficient is allowed to reflect
the $b^2 \leftrightarrow (N+1)^2/p_T^2$ dependence
that is characteristic of $k_T$ resummation, starting
from Eq.\ (\ref{powernpdy}), rather than (\ref{shapenp}).
The shape function then has
the same overall quadratic $N$-dependence as (\ref{shapenp}),
but lacks the $N$-dependence in the logarithm
and the denominator that reflects the
influence of the $K_0$ function in (\ref{fullexp}).  We thus have
\bea
\delta E_{ab}^{\mathrm np} =  \mu_0\,
\frac{C_{a}+C_{b}}{\pi}\;{(N+1)^2 \over 4p_T^2}\;
\ln\left ({2p_T \over \kappa}\right)\quad \; {\mathrm{(Fig.~2)}}\, .
\label{shapenpbonly}
\eea
Relative to Fig.~(\ref{ratioshape}), these curves show both strong
enhancements
and marked upturns toward increasing $x_T$.

\begin{figure}[h]
\centerline{\epsfig{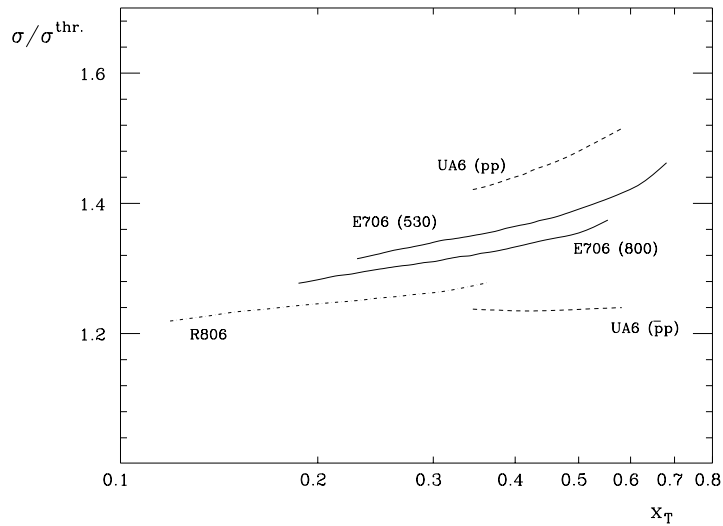}}
\caption{Same as Fig.~\ref{ratioshape}, but including the
subtracted NNLL recoil exponent (\ref{subrecoil}).
\label{recoilratios}}
\end{figure}

To complete this discussion, we consider two additional variations of
the cross sections computed with Eq.\ (\ref{shapenp}).
So far, we have ignored the term in (\ref{fullexpall})  associated
with perturbative recoil, Eq.\  (\ref{zeta2}).
It is probable that the incorporation of recoil at NNLL
would affect the values of the nonperturbative exponents.
Indeed, since both NNLL recoil and power corrections
are derived from the same starting expression,
Eq.\ (\ref{Eelaborate}), there is a serious potential for double
counting.  On the one hand,
for small values of $NQ_0/p_T$  the recoil integral in Eq.\
(\ref{Eelaborate})
is dominated by $k_T$ that are outside the soft region $k_T \le Q_0$.
On the other hand, once $N \ge p_T/Q_0$, the integration
region that gives rise to the result (\ref{zeta2}) overlaps the power
corrections almost
entirely.  Nevertheless, it is interesting to test the influence of
the corrections suggested by Eq.\ (\ref{zeta2}).
To correct for double counting, at least partially,
we use a modified estimate for recoil, which has the
property that for small $N$ it approaches (\ref{zeta2}),
while it vanishes for large $NQ_0/p_T$,
\bea
\delta \bar E_{ab,\, \mathrm rec}^{\rm SUB}
= \left( C_a+C_b \right)
{\alpha_s(4p_T^2/\bar{N}^2) \over
\pi}\ {\zeta(2)\over  2}\
\left[\, 1 - \frac{2}{\zeta(2)}\, \left(\frac{N Q_0}{2p_T} \right)^2
\ \frac{\ln \left( 1 +{\rm e}^{1/2-\gamma_E}\,
\frac{2p_T}{NQ_0}\right) }
{1+{\rm e}^{1/2-\gamma_E}\, \frac{NQ_0}{\zeta(2)p_T}}\, \right]\, .
\label{subrecoil}
\eea
For $N$ fixed, this is a leading power contribution,
with power-suppressed corrections, which, however,
conspire to cancel the leading term when $NQ_0/p_T\gg 1$.
Figure \ref{recoilratios} shows the same sets of
ratios as in Fig.~\ref{ratioshape},
including now $\delta \bar E_{ab,\, \mathrm rec}^{\rm SUB}$ in addition
to $\delta E_{ab}^{\mathrm np}$, Eq.\ (\ref{shapenp}).
To avoid double counting with the $C^{ab\to\gamma c}$ coeffcients in
the cross section, Eq.~(\ref{profilefinal}), we subtract the
leading term $(C_a+C_b)\alpha_s(4p_T^2)\zeta(2)/2\pi$ from the latter.
We see a substantial increase
compared to the pure power corrections, in addition
to a moderate slope toward large $x_T$.  We do
not take the level of this enhancement too literally,
given our rough treatment of double-counting, but
conclude that it does demonstrate the possible
importance of nonleading logarithms and
their interplay with the magnitudes of the
parameters of power corrections.

\begin{figure}[h]
\centerline{\epsfig{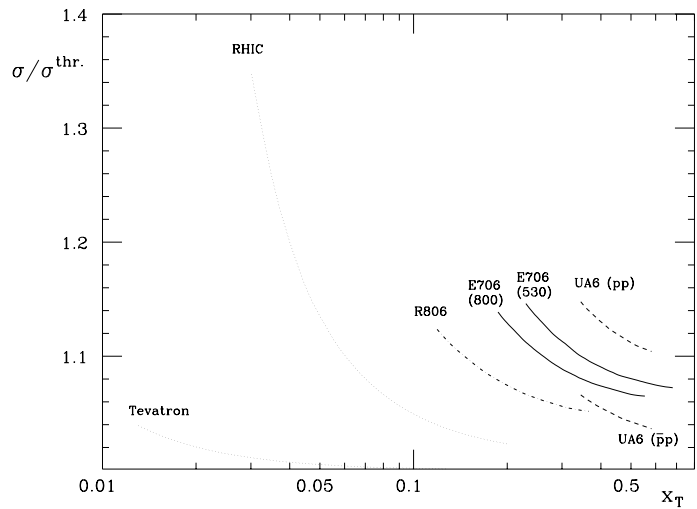}}
\caption{Same as Fig.~\ref{ratioshape}, but for a nonperturbative
function with terms nonleading in $N$ as given by Eq.~(\ref{Ebargq}).
We also show results at $\sqrt{s}=1800$~GeV and $\sqrt{s}=200$~GeV,
relevant for comparisons with Tevatron and RHIC data.
\label{barEratios}}
\end{figure}

Finally, we illustrate the possible influence of
terms that are nonleading by a power in $N$.
Although we have derived Eq.\ (\ref{profilefinal})
only for $x_T\rightarrow 1$, the form is
sufficiently general that it can be extrapolated
to any value of $x_T$.  Clearly, as we leave
the kinematic regions where large $N$ dominates,
terms that are nonleading by powers are expected to
become more and more important.
Indeed, nonleading terms may be generated from the low-scale limit
of partonic evolution.  Thinking of the pervasive  upturn of
experimental cross sections
relative to NLO noted long ago \cite{hus95},
we assume a phenomenological
parameterization for the $N$-dependent nonperturbative
exponents that behaves as $N/p_T^2$ for
large $N$, is related to the splitting functions, and
enhances the cross section at low $x_T$.
The simplest ansatz of this sort is the following modification
of the quark-gluon exponent (only),
\begin{eqnarray}
\delta {\bar {\rm E}}_{{\rm np}}^{(gq)}
&=&
\delta {\bar {\rm E}}_{\rm np}^{(g\bar q)}\ = \
\delta {\rm E}_{\rm np}^{(gq)}
+
\mu_1\, \frac{C_A}{4\pi}\; {(N+1)^2\over p_T^2}\;
\frac{1}{N-1}\, ,
\nonumber\\
\delta {\bar {\rm E}}_{\rm np}^{(q\bar q)}
&=&
\delta {\rm E}_{\rm np}^{(q\bar q)}\, .
\label{Ebargq}
\end{eqnarray}
The parameter $\mu_1$ is defined by analogy
to $\mu_0$ in (\ref{shapenp}) and (\ref{powernp}),
but does not have a direct or indirect interpretation
in terms of resummed perturbation theory.
In Fig.\ \ref{barEratios} we show the
same ratios, but now computed with the modified
shape functions (\ref{Ebargq}), choosing
$\mu_1=\mu_0=1$ GeV$^2$. These ratios
indeed show a noticeable upturn toward small $x_T$.
We observe, however, that the magnitudes of
the enhancements are nowhere near those necessary
to describe the low-$x_T$ direct photon data,
especially of E706 \cite{e706data}. Since we 
are now considering terms that are subleading at 
large $N$, we also make exploratory calculations
at higher energies, relevant to comparisons with 
the collider experiments
at Tevatron ($\sqrt{s}=1800$~GeV) and RHIC ($\sqrt{s}=200$~GeV).
As one can see, rather sharp upturns at $p_T\lesssim 5$~GeV 
are a distinct possibility here, if our ansatz in 
Eq.~(\ref{Ebargq}) is realistic.  We finally note,
without claims of physics significance, that
it is possible to provide a qualitatively
successful ($\chi^2$ per degree of freedom
approximately 1.5) ``global" fit of
direct photon data from E706, UA6, R806, and even CDF,
with the ansatz (\ref{Ebargq}), but only
for values of $\mu_1$ in the range of 10 GeV$^2$,
which implies $\mu_1C_A/4\pi \sim 2$ GeV$^2$.
The origin of such a large
scale, occuring as roughly $2\, {\rm GeV}^2/p_T^2$,
is at the least not obvious.  On the other
hand, it could be simply an artifact of using
Eq.\ (\ref{profilefinal}) outside the region
where the exponent $E(N,b)$ in (\ref{fullexp}),
evaluated at $b=-i(N+1)/p_T$, has a straightforward
interpretation.
Even more serious, however, are the potential
consequences of such corrections for pion production at collider
energies.  These issues can only be clarified by
further work.

\section{Conclusions}

We have presented an analysis of recoil and
power corrections from initial state radiation in single inclusive
direct photon cross sections at large $x_T$.
In this limit, we resum
logarithmic corrections in $N$
and simultaneously control logarithmic
and power corrections in $NQ_T/p_T$, where
$Q_T$ is a measure of partonic transverse
momentum.  Our new treatment avoids any
kinematic singularity when $Q_T$ is large.
The resulting expression is equivalent to threshold resummation
at NLL in perturbation theory, with NNLL
recoil effects. We have also shown that we may
exponentiate power corrections of
the form $NQ_T/p_T$.

In the large $x_T$ region, leading power corrections
enter in moment space as powers of $(N/p_T)^2$,
with the leading term multiplied by a logarithm of the form
$\ln(p_T/NQ_0)$.
We have observed that at large $x_T$
power corrections are suppressed relative to
expectations based on $k_T$ resummation alone.
This suppression is attributable to phase space restrictions on initial
state radiation near partonic threshold.
This result raises the possibility of a link
between a matched $k_T$ resummation
similar to that of Ref.\ \cite{fin01},
at relatively low $x_T$, and a joint
resummation at large $x_T$.
We have presented our analysis for initial state
radiation, which includes all $k_T$ dependence in
joint resummation.  A detailed discusssion including the role of final
state radiation will be given elsewhere.

Looking beyond direct photon production, we
anticipate that similar analyses may shed
light on single hadron and jet production.
A simple, but possibly significant observation
is that in single hadron cross sections,
the relevant scale for power corrections
associated with partonic threshold and
transverse momentum is $\hat{s}$, the total
partonic c.m.\ energy squared. Because $\hat{s}\ge z^{-2}(4p_T^2)$,
with $z$ the momentum fraction associated with fragmentation,
nonperturbative effects  that are inverse
powers of $\hat{s}$ are suppressed by factors of $z^2$
when expressed in terms of $p_T^2$.
Issues such as these will be relevant to
an effort to tie together perturbative
and nonperturbative effects in the full
range of inclusive hadronic reactions.

\subsection*{Acknowledgments}

The work of G.S.\ was supported in part by the National Science
Foundation, grants PHY-0098527 and PHY-0354776.
W.V.\ is grateful to RIKEN, Brookhaven National Laboratory and the U.S.
Department of Energy (contract number DE-AC02-98CH10886) for
providing the facilities essential for the completion of this work.

\section*{Appendix}

In this appendix we provide the explicit forms of the
exponents $E_{ab\to \gamma c}$, as given in~\cite{cmnov}.
According to Eq.~(\ref{splitE})
the exponent is split up into pieces associated with initial and final
state contributions. According to Eq.~(\ref{fullexp}), within our
treatment of recoil, the initial-state exponent becomes
$E_{ab,\, \mathrm{thr}}^{\rm IS}(N,p_T)+
\,E_{ab,\, \mathrm{rec}}(N,p_T)$. One has
\begin{equation}
E_{ab,\, \mathrm{thr}}^{\rm IS}(N,p_T)= \sum_{i=a,b} \left[
\frac{1}{\alpha_s (\mu^2)} h_i^{(0)} (\lambda) +
h_i^{(1)} (\lambda,2p_T,\mu, \mu_F) \right]  \;  ,
\end{equation}
where
\begin{eqnarray}
h_i^{(0)} (\lambda) &=& \frac{A_i^{(1)}}{2\pi b_0^2}
\left[ 2 \lambda + (1 - 2 \lambda) \ln(1-2 \lambda) \right]\, ,
\label{hsubadef0}
\end{eqnarray}
and
\begin{eqnarray}
h_i^{(1)} (\lambda,2p_T,\mu,\mu_F) &=&
\frac{A_i^{(1)} b_1}{2\pi b_0^3} \left[ \frac{1}{2} \ln^2 (1-2 \lambda)
+
2 \lambda + \ln(1-2 \lambda)\right]  \\
&+& \frac{1}{2\pi b_0} \left( - \frac{A_i^{(2)}}{\pi b_0} +
A_i^{(1)} \ln \left( \frac{4 p_T^2}{\mu^2} \right) \right)
\left[ 2 \lambda + \ln(1-2 \lambda) \right]-
\frac{A_i^{(1)}}{\pi b_0} \lambda \ln \left( \frac{4 p_T^2}{\mu_F^2}
\right) \; .\nonumber
\label{hsubadef}
\end{eqnarray}
For completeness, we have distinguished between the renormalization
scale $\mu$ and the factorization scale $\mu_F$.
The $A_i^{(1)}$ are as in Eq.~(\ref{explicitA}), and we have defined
\begin{eqnarray}
\lambda &=& b_0 \alpha_s (\mu^2)  \ln \bar N \; , \nonumber \\
b_0 &=& \frac{11 C_A - 4 T_R N_F}{12 \pi}\, , \nonumber \\
b_1 &=& \frac{17 C_A^2-10 C_A T_R N_F-6 C_F T_R N_F}{24 \pi^2}\, .
\end{eqnarray}
For the NNLL exponent $E_{ab,\, \mathrm{rec}}(N,p_T)$ we obtain
\begin{equation}
E_{ab,\, \mathrm rec}^{\rm NNLL} = \left( C_a+C_b \right)
{\alpha_s(\mu^2) \over
2\pi(1-2 \lambda)}\ \zeta(2)\; .
\end{equation}
The exponent for the final state reads:
\begin{equation}
E_{abc}^{\rm FS}(N,2p_T,\mu)=\frac{1}{\alpha_s (\mu^2)} f^{(0)}_c
(\lambda) +
f^{(1)}_c (\lambda,2p_T,\mu)+g^{(1)}_{abc} (\lambda) \, ,
\end{equation}
with
\begin{eqnarray}
f^{(0)}_a (\lambda) &=& 2 h^{(0)}_a (\lambda/2) - h^{(0)}_a
(\lambda) \; ,\\
f^{(1)}_a (\lambda,2p_T,\mu) &=&
2 h^{(1)}_a (\lambda/2,2p_T,\mu,2p_T) -
h^{(1)}_a (\lambda,2p_T,\mu,2p_T) \nonumber \\
&+& \frac{A_a^{(1)} \ln 2}{\pi b_0}
\left( \ln(1-2 \lambda) - \ln(1-\lambda) \right)
- \frac{B_a^{(1)}}{\pi b_0}  \ln(1-\lambda) \;  , \\
g^{(1)}_{q\bar{q}g} (\lambda) &=& -\frac{C_A}{\pi b_0}
\ln(1-2 \lambda) \ln 2  \; ,\\
g^{(1)}_{qgq} (\lambda) &=& -\frac{C_F}{\pi b_0} \ln(1-2 \lambda)
\ln 2 \, .
\end{eqnarray}
Here,
\begin{eqnarray}
B^{(1)}_q = \frac{3}{4} C_F \; , \quad
B^{(1)}_g= {\beta_0\over 4} \, .
\label{explicitB}
\end{eqnarray}
Finally, the coefficients $C^{ab\to \gamma c}$ of Eq.~(\ref{1pIresumE})
read~\cite{cmnov}:
\begin{eqnarray}
C^{q{\bar q}\to \gamma g} &=&
1+\frac{\alpha_s}{\pi}\Bigg[
- \, \frac{1}{2} (2 C_F - C_A) \ln 2
+ \frac{1}{2} \,K - K_q
+ 2 \zeta(2) \Bigg( 2 C_F - \frac{1}{2} C_A \Bigg)\nonumber \\
&&+ \frac{5}{4} (2 C_F - C_A) \ln^2 2 +
\frac{3}{2} C_F  \ln \frac{2 p_T^2}{\mu_F^2}
- \pi b_0 \ln \frac{2 p_T^2}{\mu^2} \Bigg] \;\;, \\
C^{qg\to \gamma q} &=&
1+\frac{\alpha_s}{\pi}\Bigg[
- \, \frac{1}{10} ( C_F - 2 C_A) \ln 2 - \frac{1}{2} \,K_q
+ \frac{\zeta(2)}{10} \Bigg( 2 C_F + 19 C_A \Bigg)
+ \frac{1}{2} C_F  \ln^2 2 \nonumber\\
&&+ \Bigl(  \frac{3}{4} C_F + \pi b_0 \Bigg)
\ln \frac{2 p_T^2}{\mu_F^2}
- \pi b_0 \ln \frac{2 p_T^2}{\mu^2} \Bigg] \;\;,
\end{eqnarray}
where
\begin{eqnarray}
\label{kcoeffq}
K &=& C_A \left( \frac{67}{18} - \frac{\pi^2}{6} \right)
- \frac{10}{9} T_R N_f \;\;, \nonumber \\
K_q &=& \left( \frac{7}{2} - \frac{\pi^2}{6} \right) \,C_F \;\;.
\end{eqnarray}

\end{document}